\documentclass{iau-JDSS}
\usepackage{graphicx}

\title[JD 15.~~Turbulent magnetic reconnection in 2D and 3D] 
{2D and 3D turbulent magnetic reconnection}

\author[A. Lazarian et al.]   
{A. Lazarian$^1$,
 G. Kowal$^1$,
 E. Vishniac$^2$,
 K. Kulpa-Dubel$^3$,\and
 K. Otmianowska-Mazur$^3$
}

\affiliation{
$^1${University of Wisconsin-Madison\\ {\tt lazarian@astro.wisc.edu}} \\[\affilskip]
$^2${McMaster University}\\[\affilskip]
$^3${Kracow University}\\[\affilskip]
}

\pubyear{2010}
\volume{15}
\pagerange{31--32}
\date{May}
\jname{IAU Highlights of Astronomy, Volume 15}
\editors{Ian F. Corbett, ed.}

\begin{document}

\maketitle

\begin{abstract}
Magnetic field embedded in a perfectly conducting fluid preserves its topology for all time.  Although ionized astrophysical objects, like stars and galactic disks, are almost perfectly conducting, they show indications of changes in topology, `magnetic reconnection’, on dynamical time scales.  Reconnection can be observed directly in the solar corona, but can also be inferred from the existence of large scale dynamo activity inside stellar interiors.  Solar flares and gamma ray busts are usually associated with magnetic reconnection. Previous work has concentrated on showing how reconnection can be rapid in plasmas with very small collision rates.  Here we present numerical evidence, based on three dimensional simulations, that reconnection in a turbulent fluid occurs at a speed comparable to the rms velocity of the turbulence, regardless of the value of the resistivity.  In particular, this is true for turbulent pressures much weaker than the magnetic field pressure so that the magnetic field lines are only slightly bent by the turbulence.  These results are consistent with the proposal by Lazarian \& Vishniac (1999) that reconnection is controlled by the stochastic diffusion of magnetic field lines, which produces a broad outflow of plasma from the reconnection zone.  This work implies that reconnection in a turbulent fluid typically takes place in approximately a single eddy turnover time, with broad implications for dynamo activity and particle acceleration throughout the universe. In contrast, the reconnection in 2D configurations in the presence of turbulence depends on resistivity, i.e. is slow. 

\end{abstract}

\noindent
{\bf Reconnection in 3 dimensions versus 2 dimensions}

Turbulence is ubiquitous in magnetized astrophysical fluids and it makes magnetic reconnection fast according to the model by Lazarian \& Vishniac (1999, henceforth LV99). The LV99 model naturally generalizes Sweet-Parker model for the case of turbulent field lines (Fig. 1). Unlike earlier attempts to invoke turbulence into reconnection, the LV99 model provided testable predictions of how reconnection rate changes with the power and scale of the turbulence. 

The testing of the model in Kowal et al. (2009) were found consistent with the prediction of the LV99 model. For instance, Fig. 2a shows
that magnetic reconnection is independent of resistivity. Taking into account the differences of the Lundquist numbers of numerical simulations and astrophysical fluids this would not constitute a solid prove of the reconnection being fast. However, Kowal et al. (1999) also successfully tested the scaling of the reconnection rates predicted by LV99. This allows us to claim the success of the LV99 model of fast reconnection. 

An explanatory note is also due. The LV99 model does not require turbulence being strong, i.e. it does not require strong bending of magnetic field lines by turbulence. The testing of the model in Kowal et al. (2009) were performed for subAlfvenic driving. At the similar circumstances, turbulence in 2D, which was the focus of the research prior to LV99 work does not show fast reconnection Fig. 2b. (Kulpa-Dubel et al. 2009). This confirms the assesment in LV99 that the turbulent reconnection is only fast in 3D. 

\begin{figure*}[!t]
\begin{center}
  \includegraphics[width=0.45\columnwidth]{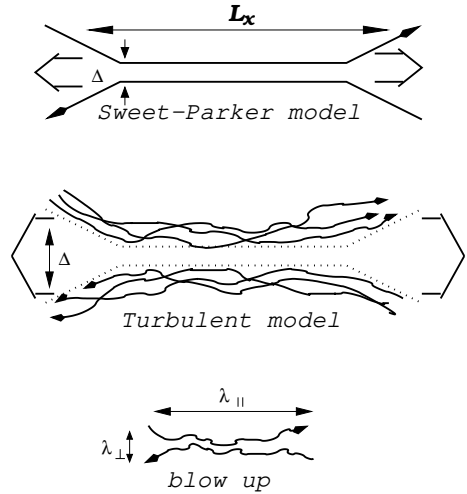}%
\end{center}
  \caption{{\it Upper plot}: 
Sweet-Parker model of reconnection. The outflow
is limited by a thin slot $\Delta$, which is determined by Ohmic 
diffusivity. The other scale is an astrophysical scale $L\gg \Delta$.
{\it Middle plot}: Reconnection of weakly stochastic magnetic field according to 
LV99. The model that accounts for the stochasticity
of magnetic field lines. The outflow is limited by the diffusion of
magnetic field lines, which depends on field line stochasticity.
{\it Low plot}: An individual small scale reconnection region. The
reconnection over small patches of magnetic field determines the local
reconnection rate. The global reconnection rate is substantially larger
as many independent patches come together. From Lazarian et al. (2004).}
  \label{fig:1}
\end{figure*}

\begin{figure*}
\begin{center}
  \includegraphics[width=0.40 \columnwidth]{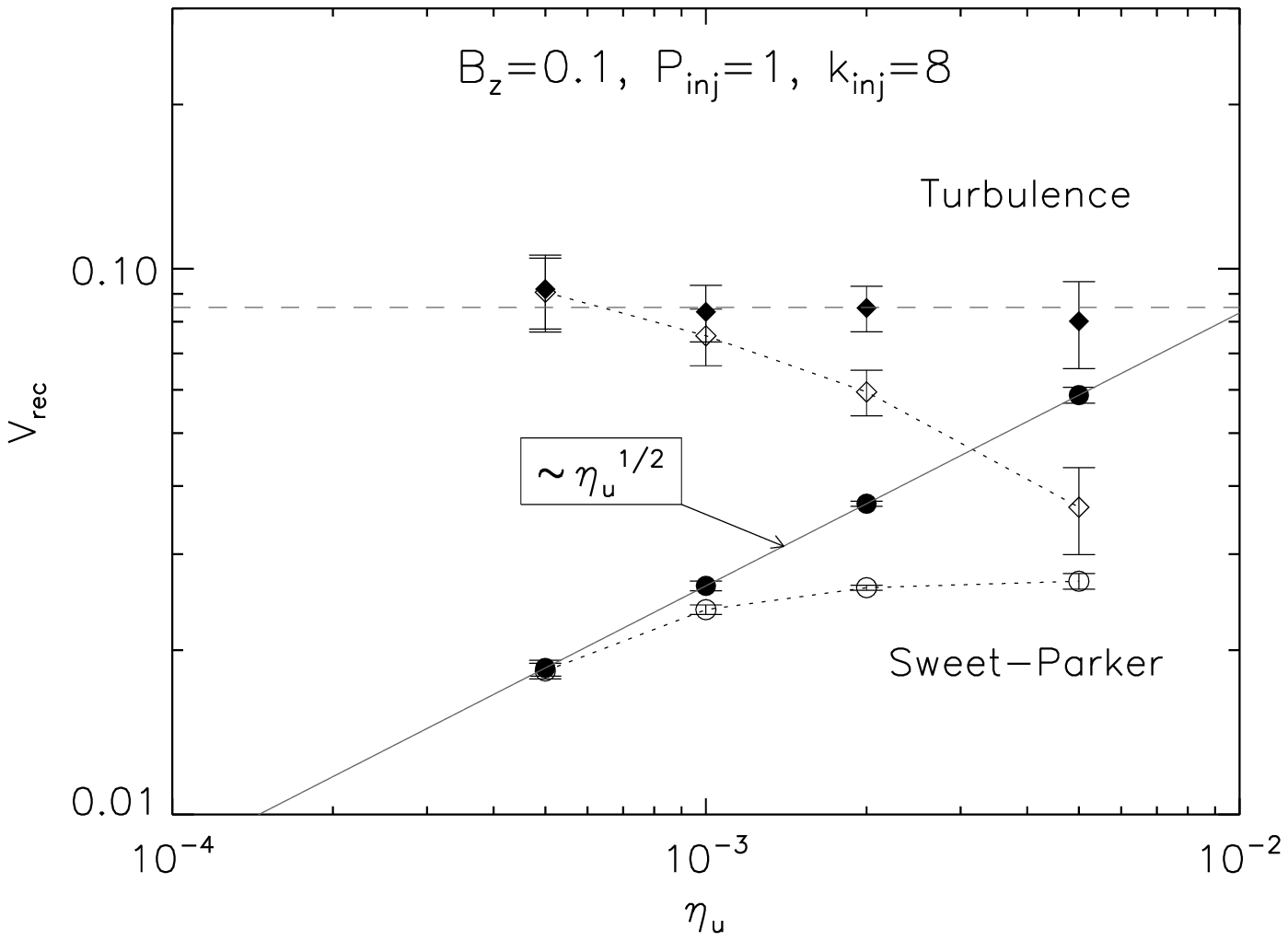}%
  \hspace*{\columnsep}%
  \includegraphics[width=0.40 \columnwidth]{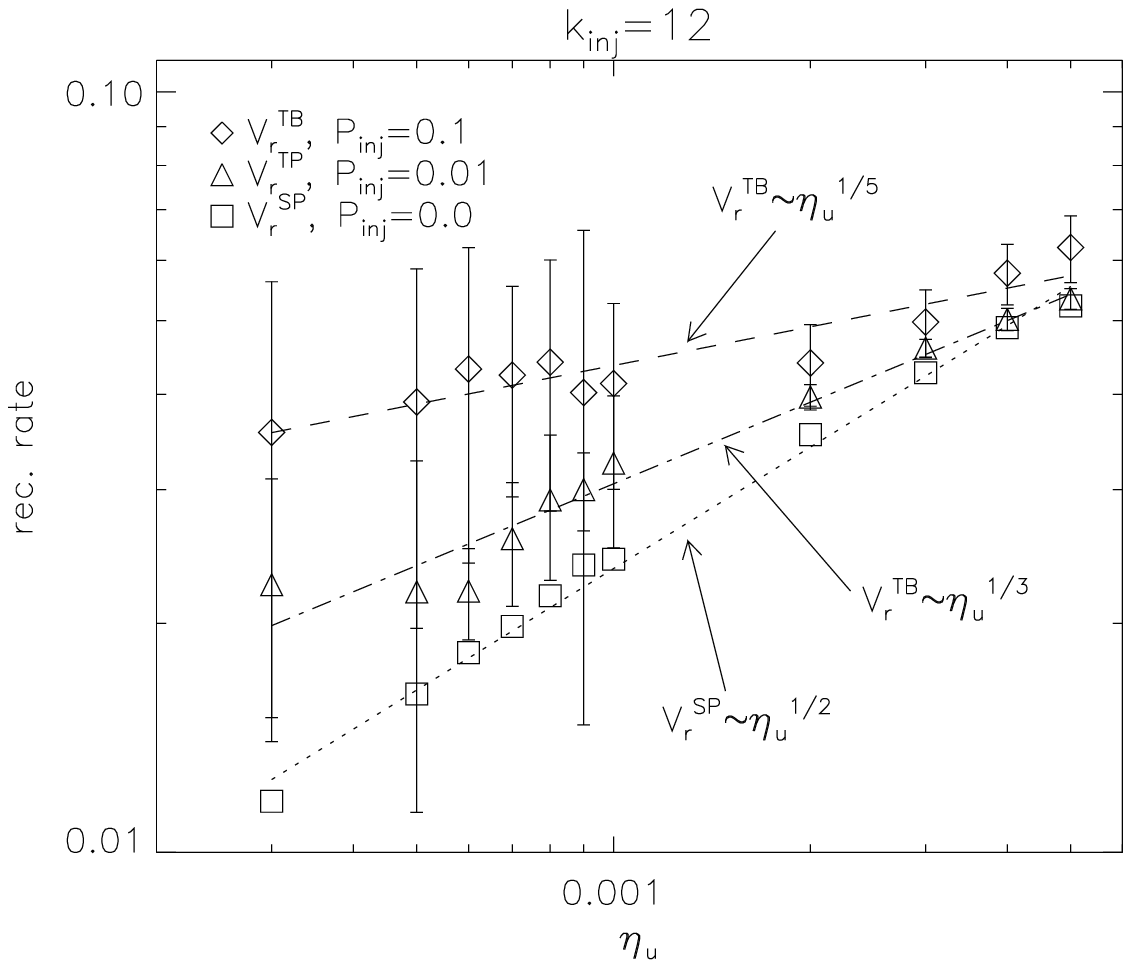}
\end{center}
  \caption{{\it Left panel}: Reconnection in 3D is fast, i.e. independent of resistivity, when turbulence is present. {\it Right panel}: Reconnection in 2D is slow in the presence of turbulence, i.e. it depends on resistivity. Even weak dependence of resistivity makes reconnection negligible in astrophysical consitions. }
  \label{fig:2}
\end{figure*}

\end{document}